\documentclass[numbers,webpdf]{ima-authoring-template}%
\usepackage{cleveref}

\graphicspath{{Fig/}}

\newcommand{\curriculum}{programme}

\newcommand{\changed}[1]{{\color{blue} #1}}
\renewcommand{\changed}[1]{{ #1}}

\theoremstyle{thmstyletwo}%
\numberwithin{equation}{section}

\begin{document}

\DOI{DOI HERE}
\copyrightyear{2025}
\vol{00}
\pubyear{2025}
\access{Advance Access Publication Date: Day Month Year}
\appnotes{Paper}
\copyrightstatement{Published by Oxford University Press on behalf of the Institute of Mathematics and its Applications. All rights reserved.}
\firstpage{1}


\title[GenAI performance in undergraduate mathematics]{Generative AI performance in core undergraduate mathematics: a curriculum-level case study}

\author{Benjamin J. Walker\ORCID{0000-0003-0853-267X}*, Nikoleta Kalaydzhieva\ORCID{0000-0002-9393-283X}, Beatriz Navarro Lameda\ORCID{0000-0002-9092-2492} and Ruth A. Reynolds\ORCID{0000-0001-9738-9013} 
\address{\orgdiv{Department of Mathematics}, \orgname{University College London}, \orgaddress{\street{Gordon Street}, \postcode{WC1H 0AY}, \state{London}, \country{UK}}}}

\authormark{Benjamin J. Walker et al.}

\corresp[*]{Corresponding author: \href{email:benjamin.walker@ucl.ac.uk}{benjamin.walker@ucl.ac.uk}}

\received{Date}{0}{Year}
\revised{Date}{0}{Year}
\accepted{Date}{0}{Year}


\abstract{Generative artificial intelligence (GenAI) tools such as OpenAI’s ChatGPT are transforming the educational landscape, prompting reconsideration of traditional assessment practices. In parallel, universities are exploring alternatives to in-person, closed-book examinations, raising concerns about academic integrity and pedagogical alignment in uninvigilated settings. This study systematically investigates the performance of GenAI on typical mathematics questions from across a first-year mathematics curriculum. Adopting an empirical approach and utilising current examination questions as a proxy for course content, we generate, transcribe, and blind-mark GenAI submissions to eight undergraduate mathematics assessments, spanning the entirety of the first-year curriculum. By combining independent GenAI responses to individual questions, we enable a meaningful evaluation of GenAI performance, both at the level of modules and across the first-year curriculum. We find that GenAI attainment is at the level of a first-class degree, though current performance can vary between modules. Further, we find that GenAI performance is remarkably consistent when viewed across the entire curriculum, significantly more so than that of students in invigilated examinations. Our findings evidence the pressing need for redesigning assessments in mathematics in the era of generative artificial intelligence.}
\keywords{Generative artificial intelligence (GenAI); Mathematics assessment; Open-book examination.}


\maketitle

\section{Introduction}
Generative artificial intelligence (GenAI) is increasingly impacting education \cite{freeman2025student,UKGovReport}. Here and throughout, we use GenAI to broadly refer to any system that is capable of producing text, images, and other media in response to user prompts. Such systems, including ChatGPT \cite{chatgpt}, Gemini \cite{google_gemini}, Perplexity \cite{perplexity}, and countless others, often generate content with human-like fluency and interactivity, features that have likely helped drive their popularity and widespread adoption.

GenAI can be found in one form or another in school classrooms \cite{UKGovReport}, undergraduate courses \cite{freeman2025student}, and international competitions \cite{Castelvecchi2025}. Overall, this rapid proliferation presents both an opportunity for innovation and the pressing need for adaptation \cite{Sallai2024}. Whilst early discussions focused on the immediate potential for disruption to academic integrity, the conversation is shifting towards understanding the capabilities of these tools to inform long-term curriculum design \cite{Qian2025}. Use cases include GenAI as tools for personalised study support and as sources of sample material for critical appraisal by students \cite{freeman2025student,Baytas2025}. Other examples at this rapidly evolving frontier include GenAI-assisted assessments \cite{perdikoulias2025generativeaimultiplechoice,caraeni2024evaluating,Zubaer2025,Isley2025,Haddley2024}, learning materials \cite{https://doi.org/10.1111/ssm.18356,Isley2025}, and even lesson planning and curricula \cite{Grove2025,bektik2024ai,Pepin2025,Gurl2025}. There are also widespread anecdotal reports of students using GenAI for \changed{a range of }assessments, supported by a recent study in which 88\% of students in their sample made use of such tools when completing assignments \cite{freeman2025student}.

Associated with this broadening adoption of GenAI is an increasing need to rigorously evaluate its capabilities. As is to be expected for an emerging technology, its abilities are evolving rapidly; whilst early evaluations in mathematics observed frequent errors in even simple tasks \cite{yen2023questionsconcerninguselarge,barana2023fostering,dao2023investigating}, more recent performance demonstrates proficiency in ever more complex scenarios \cite{Fang2025,Wang2025llmMathProblemSolving_v1,Frieder,moon2025performance}. GenAI has also had notable recent success in advanced, competition-style mathematics \cite{Castelvecchi2025}, and its capabilities have been assessed in a range of subjects and contexts \cite{Coello2024,Scarfe2024,Benam2024,Bohlmann2024,wang-etal-2025-llms-perform,Zubaer2025,Katz2024,Kung2023,Chongchitnan2025}. However, despite the growing body of literature making use of and evaluating GenAI, to the best of our knowledge there remains a gap in understanding how GenAI performs holistically across a standard UK undergraduate programme in mathematics.

In this case study, it is our aim to provide a systematically constructed basis of evidence for the performance of GenAI across a core portion of mathematics curricula in higher education institutions. This evidence is intended as a basis for institutional decision making, enabling discussions about GenAI attainment and capability in departments and universities to move beyond anecdotal reports. Stated more precisely, we seek to address the following research question:
\begin{quote}
    To what extent does GenAI demonstrate mastery of first-year undergraduate mathematics learning outcomes when evaluated against the breadth of a standard university curriculum?
\end{quote}
Drawing inspiration from a similar study in psychology \cite{Scarfe2024}, we adopt a rigorous, empirical approach by generating, transcribing, and blind-marking GenAI responses to real assessments. \changed{We focus on the performance of a popular GenAI tool with limited prompting\footnote{In particular, we do not provide specific course materials as part of the prompts used in this study.}; further details can be found in \cref{sec: methods: produce and mark}.} \changed{We utilise the full range of first-year undergraduate mathematics examinations at a Russell Group university as a proxy for demonstrating mastery, noting that closed-book examinations are typically given a high weighting in UK universities \cite{Iannone2022}. This necessarily limits our definition of mastery to performance in written examinations alone, which facilitates quantitative comparison but falls short of a traditional definition of mastery \cite{bloom1968learning}, as would typically be measured with a broader assessment diet.} By combining independent responses to individual questions, we enable a meaningful evaluation of GenAI performance, both at the level of modules and across the curriculum as a whole.

The remainder of this case study is structured as follows. We outline our approach in \cref{sec: methods} before presenting our results in \cref{sec: results}. We conclude by summarising our findings and discussing them in context in \cref{sec: discussion}.

\section{Methods}\label{sec: methods}
We conducted a systematic evaluation of GenAI across a first-year mathematics curriculum. To do this, we utilised the eight written examinations from academic year 2024--2025 at a Russell Group university as a proxy for the mathematics curriculum. To provide a baseline for performance, we compared GenAI and student attainment across these assessments.

The selected examinations formed the primary mode of assessment for the eight core modules in the first year of the BSc/MSci Mathematics programmes at the university, and are expected to be typical of the format and content found in undergraduate mathematics programmes in the UK. Detailed descriptions of the modules are provided in \cref{app: modules}. Examinations were delivered to students as closed-book assessments, in that notes and reference materials were not available to students. Each examination consisted of four or five multipart questions totalling 100 marks, which were moderated as per standard practice at the university and not subject to post-hoc scaling.

In brief, multiple GenAI responses to all 33 questions from the 2024--2025 exam period were generated, transcribed, and blind-marked along with the scripts of the cohort of students sitting these examinations. We then combined GenAI responses to simulate a large number of possible exam entries, facilitating comparisons at the \emph{module level}. Finally, to enable a meaningful comparison at the \emph{\curriculum{} level}, we combined scores across the eight modules, generating an average mark for each student and each simulated set of GenAI responses. We describe this process in detail below and report our findings in \cref{sec: results}. We focus exclusively on GenAI attainment, and do not evaluate the detection of GenAI submissions. Indeed, we go to significant lengths to mask our GenAI submissions in order to facilitate unbiased grading. This contrasts with the work of \citet{Scarfe2024}, which considered both attainment and detection in uninvigilated assessments in psychology.

\subsection{Producing and marking GenAI exam submissions}\label{sec: methods: produce and mark}
In this study, we used OpenAI's popular GenAI platform ChatGPT to generate responses to exam questions. Before submission to ChatGPT, each question was converted to plain text format, making use of standard \LaTeX ~commands for mathematical notation. One question included reference to a diagram; this diagram was replaced with a detailed description and the apparent comprehension of ChatGPT was not adversely affected. In order to elicit responses in an appropriate format and style, the following prompt was prepended to each question:
\begin{quote}
    You are a mathematics undergraduate student. Respond as if this were an examination at a UK university. Be precise, thorough and concise.
\end{quote}
\changed{Though seemingly innocuous, we remark that our use of this prompt may have had unintended and uncertain effects on the outputs produced. For instance, it is not clear if being described as a mathematics undergraduate student will help or hinder GenAI performance; similar such effects on the variability of responses are also uncertain. In this study, we observed that this prompt routinely led to responses that were in an appropriate style, of an appropriate length, and largely at the level expected of undergraduate students, and defer a detailed investigation of prompt effectiveness to future study.}

With this directive, plain text questions (with all subparts included) were submitted via OpenAI's API \cite{openai_api_reference} to the model \texttt{chatgpt-4o-latest} in March 2025. This model corresponded to that used in the ChatGPT web interface at the time of submission. Eight responses were requested for each question in order to capture some of the variation in GenAI responses; the first five complete\footnote{Rarely, GenAI responses ended abruptly before providing a complete solution. Such responses were discarded.} responses were used in order to balance the corresponding additional marking load. We emphasise that responses generated in this way are purportedly independent of one another \changed{in a statistical sense, in that the generation of a response does not influence subsequent responses}. Responses invariably utilised valid \LaTeX~syntax and were compiled into individual PDFs.

We note that this approach is highly unlikely to yield the highest quality output from GenAI; given refined prompting and the submission of course materials (such as lecture notes), we expect that attainment could be drastically improved. As such, the performance of GenAI reported in this study should only be considered as a crude lower bound. As we will note in \cref{sec: discussion}, this has little impact on our conclusions.

GenAI responses were transcribed by hand into 40 exam scripts during the main exam period, then scanned and submitted to the digital marking platform used at the institution. By design, GenAI scripts were practically indistinguishable from scripts produced by students. These scripts were marked alongside the student's scripts; to the best of our knowledge, markers were not aware of the presence of the GenAI scripts when marking began.

\subsection{Comparing student and GenAI cohorts}
Following marking, GenAI scripts were separated and the marks for individual questions recorded\changed{, resulting in 165 marked responses across the 33 questions}. This immediately enabled question-level comparison of students and GenAI, though with modest GenAI sample size. To enable module-level comparison, we exploited the \changed{statistical} independence of GenAI responses to individual questions by combining them to yield estimates for performance on each module. The same reasoning applied at the \curriculum{} level: we averaged the module-level GenAI performance across the eight exams to estimate the attainment across the first-year programme.

\begin{figure}[H]
    \centering
    \includegraphics[width=0.6\linewidth]{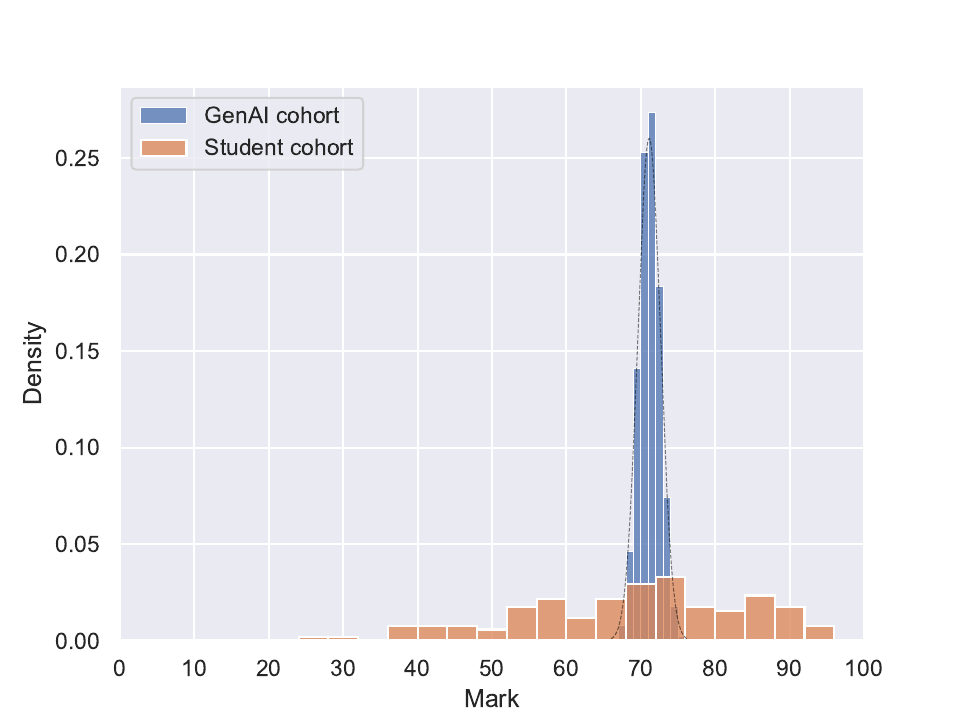}
    \caption{Comparing GenAI and student performance across the curriculum. We present the averaged attainment of both GenAI and student cohorts across all eight modules in the first-year programme, with the GenAI and student cohorts shown in blue and orange, respectively. Visually, it is clear that GenAI can attain similar marks to students but is much more consistent overall. In terms of mean marks, GenAI appeared to marginally outperform students (71.1 vs 69.1, respectively), though this difference was not statistically significant (two-sided t-test, $p = 0.13$). The overall variance of the GenAI cohort was significantly lower than that of the student body (standard deviations of 1.53 vs 15.1, respectively; Levene's test, $p < 10^{-15}$). Unbiased estimates of the mean and variance of GenAI attainment were combined to give a normal probability density estimate, shown here as a black dashed curve and approximately indistinguishable from the empirical distribution shown in blue.}
    \label{fig: programme}
\end{figure}

We used two approaches to combine questions and modules: resampling and direct estimation. In the former, we generated a large sample of GenAI marks for each module by taking all possible combinations of the marks attained for individual questions. The resulting distributions are shown as histograms \changed{in \cref{fig: programme}}. These large samples were themselves combined across the eight modules to give a \curriculum{}-level sample of GenAI performance, with each module weighted equally. This yielded a combinatorially large number of sampled GenAI marks at the \curriculum{} level, which we randomly sampled from to generate \cref{fig: programme}. As in bootstrapping, following this procedure and estimating the variance systematically underestimated the population variance, in this case by a factor of $5/4$. In practice, we observed that this made no appreciable difference to the histograms of performance at the module and \curriculum{} levels. For completeness, we overlay the biased, resampled histograms \changed{in \cref{fig: programme}} with unbiased normal estimates, obtained as described below. We remark that the estimator of the mean is unbiased.

In parallel, we adopted the simpler approach of directly estimating the population mean and variance of GenAI responses to each question. Given the \changed{statistical} independence of the questions from one another, these estimates were summed to give unbiased estimates for the mean and variance of GenAI attainment at the module level. Finally, the module-level estimates were averaged (noting the quadratic scaling of variance under linear transformations) to yield unbiased estimates for the mean and variance of GenAI performance at the \curriculum{} level. Notably, under reasonable but technical assumptions, aggregated attainment (both at the module and \curriculum{} levels) is expected to be approximately normal in distribution. Thus, the distributions can be well described by their mean and variance, as evidenced by the agreement shown in \cref{fig: programme,fig: modules}.

In each of our comparisons, we report the results of the 128 students who participated in all eight exams, having excluded any students with one or more absences. Marks are \changed{shown} for each question, each module, and the average of the eight modules for each student. Note that we did not assume independence of a student's marks at the question or module level; indeed, student performance was highly correlated across modules. Population means and variances of awarded marks were computed for each question, each module, and the programme.

\section{Results}\label{sec: results}

\subsection{GenAI achieves comparable mean performance to students across the curriculum, with markedly lower variance}
Following the approach outlined above, we obtained distributions of student and GenAI attainment across the first-year curriculum, as shown in \cref{fig: programme} \changed{for the 128 students and 165 GenAI responses to individual questions}. Overall, we observed an average GenAI mark of 71.1 out of 100, equivalent to a low first-class degree and statistically indistinguishable\footnote{Two-sided t-test comparing the student cohort to the estimated mean GenAI attainment, $p=0.13$.} from the student cohort’s mean (69.1).

Whilst the average performances of students and GenAI were similar, the distributions of attainment differ substantially. As is evident from \cref{fig: programme}, GenAI performed markedly more consistently than students, with estimated standard deviations differing by an order of magnitude (1.5 and 15.1, respectively). This significant difference\footnote{Levene's test, $p < 10^{-15}$.} in spread can largely be attributed to the independence of individual GenAI responses to questions; expectedly and contrastingly, student responses were highly correlated. Nonetheless, we observed that GenAI performance on individual questions was notably less variable than that of the student cohort, as presented in detail in \cref{app: questions}.

\subsection{Module performance highlights topic-specific variability in GenAI attainment}

\begin{figure}[H]
    \centering
    \includegraphics[width=0.8\linewidth]{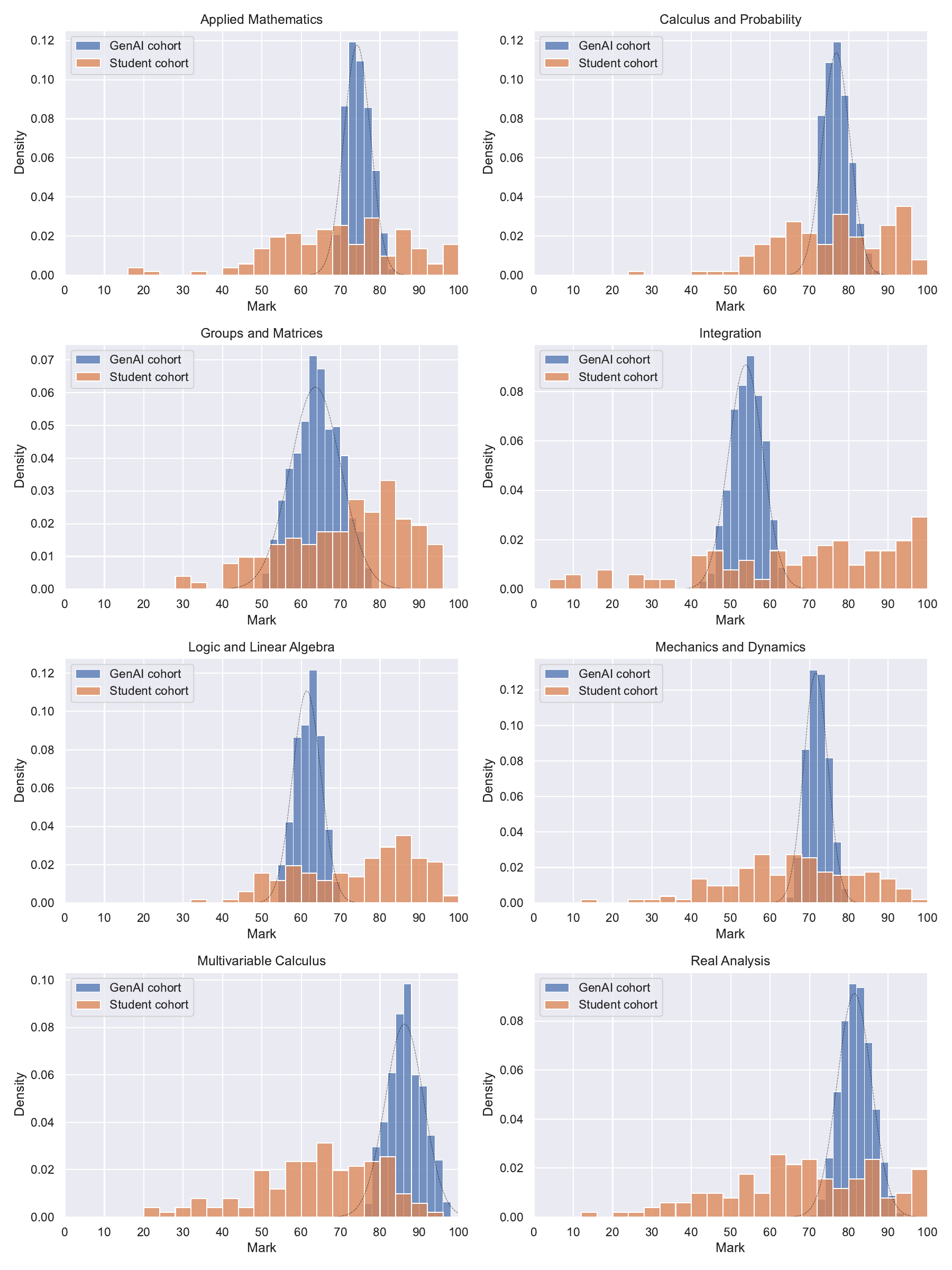}
    \caption{Comparing GenAI and student performance in each of the eight modules of the first-year programme of study, with the GenAI and student cohorts shown in blue and orange, respectively. Whilst the curriculum-level performance of GenAI shown in \cref{fig: programme} is strong, module-level performance is highly variable. Within each module, GenAI attainment was significantly less variable than that of the student cohort. Unbiased estimates of the mean and variance of GenAI attainment were combined to give a normal probability density estimate, shown here as black dashed curves and approximately indistinguishable from the empirical distributions shown in blue.}
    \label{fig: modules}
\end{figure}

\begin{table*}
\caption{Summary statistics for GenAI and student submissions across the curriculum.\label{tab: marks}}
\tabcolsep=0pt
\begin{tabular*}{\textwidth}{@{\extracolsep{\fill}}lcccc@{\extracolsep{\fill}}}
\toprule%
& \multicolumn{2}{@{}c@{}}{GenAI cohort} & \multicolumn{2}{@{}c@{}}{Student cohort} \\
\cline{2-3}\cline{4-5}%
Module & Mean & Standard deviation & Mean & Standard deviation \\
\midrule
     Applied Mathematics & 74.2 & 3.4 & 69.4 & 16.8 \\
Calculus and Probability & 76.8 & 3.5 & 75.4 & 14.2 \\
     Groups and Matrices & 63.6 & 6.5 & 70.6 & 15.4 \\
             Integration & 53.8 & 4.4 & 66.8 & 24.8 \\
Logic and Linear Algebra & 61.4 & 3.6 & 74.0 & 15.0 \\
  Mechanics and Dynamics & 71.6 & 3.1 & 65.5 & 16.4 \\
  Multivariable Calculus & 86.2 & 4.9 & 63.7 & 15.5 \\
           Real Analysis & 81.4 & 4.4 & 67.2 & 19.0 \\
\midrule
Overall & 71.1 & 1.5 & 69.1 & 15.1\\
\botrule
\end{tabular*}
\begin{tablenotes}%
\item Brief module descriptions are provided in \cref{app: modules}. When estimating standard deviation, GenAI marks are considered to be samples from a larger population; student marks are considered to represent an entire population.
\end{tablenotes}
\end{table*}

Overall, the performance of GenAI across the curriculum revealed consistently high attainment, comparable to or exceeding that of the cohort of students. However, performance between modules varied significantly. In \cref{fig: modules,tab: marks}, we present a detailed, module-level comparison of attainment, revealing clear areas of relative GenAI strength and weakness.

For instance, in Multivariable Calculus, GenAI substantially outperformed students. This is partially attributable to the presence of a programming question \changed{(Question 1, see \cref{fig: questions 2})}, to which GenAI reliably provided essentially perfect solutions. Given the prevalence of GenAI tools in software engineering, its strong performance is perhaps unsurprising. \changed{We remark, however, that GenAI performance on a non-programming question (Question 3, see \changed{\cref{fig: questions 2}}) in the same module was higher.}

In contrast, GenAI's attainment in Integration was relatively low. Detailed inspection of the marked scripts revealed the source of this seemingly systematic error: GenAI had assumed results that were beyond the scope of the course. Whilst we believe the content of the examination to be typical of first-year undergraduate mathematics, the results used in the GenAI solutions are typically taught in the third year of a mathematics undergraduate programme. No module-specific information (such as level or lecture notes) had been provided to the tool. Hence, this result should not be unexpected, and highlights the potential importance of providing sufficient context to GenAI tools. However, we note that this did not seem to be the case in other parts of the curriculum, even in Real Analysis.

\subsection{Question-level analysis identifies themes in GenAI performance}

\begin{table*}
\caption{GenAI performance grouped using the taxonomy of \citet{SmithTaxonomy}, reported as percentage attainment across the questions in each taxon. \label{tab: classification}}
\tabcolsep=0pt
\begin{tabular*}{\textwidth}{@{\extracolsep{\fill}}lcclc@{\extracolsep{\fill}}}
\toprule%
& \multicolumn{3}{@{}c@{}}{GenAI cohort (\%)} & \\
\cline{2-4}%
Classification & Mean & Standard deviation & Mode (count) & Number of questions \\
\midrule
                Applications in new situations & 67 & 25 &  100 (51) & 33 \\
                         Comprehension & 82 & 27 &  100 (45) & 13 \\
                     Factual knowledge & 82 & 28 & 100 (120) & 35 \\
Implications, conjectures, comparisons & 69 & 27 &    33 (5) &  3 \\
                  Information transfer & 69 & 19 &  100 (18) & 10 \\
           Justifying and interpreting & 70 & 37 &  100 (27) & 14 \\
             Routine use of procedures & 76 & 27 & 100 (108) & 42 \\
\botrule
\end{tabular*}
\begin{tablenotes}%
\item Questions have been subdivided into subparts so as to better align with a single taxon. Means, standard deviations, and modes (including the number of responses attaining the mode) are reported for each taxon, along with the number of questions/subparts in that taxon. Note that the total number of responses is five times the number of questions/subparts in any given taxon.
\end{tablenotes}
\end{table*}

In addition to module-level differences, attainment varied between individual questions. \changed{Detailed distributions showing GenAI and student performance on each question can be found in \cref{app: questions}. 

In order to systematically assess the performance of GenAI on various types of questions, we classified the questions in the examinations following the taxonomy of \citet{SmithTaxonomy}, subdividing questions into subparts so as to align them with unique taxa. In \cref{tab: classification}, we report the mean and modal percentage attainment of GenAI on questions in each taxon, along with the standard deviation of the mean attainment and the number of questions in each taxon. Questions were largely classified as \emph{Routine use of procedures}, \emph{Factual knowledge}, or \emph{Applications in new situations}; only three subparts were classified in the \emph{Implications, conjectures, comparisons} taxon. Of note, the modal value in all but one taxon was 100\% attainment, despite somewhat variable means and appreciable standard deviations. Alongside highlighting a bias towards high-scoring responses, these observations entail that the summary statistics of \cref{tab: classification} should be treated with caution: excluding peaks at 100\%, the underlying distributions of attainment are approximately uniform. We refer the reader to \cref{fig: taxonomy} in \cref{app: questions} for plots of the distributions, along with a graphical summary of the distribution of question types.

With these caveats in mind and acknowledging the large standard deviations reported in \cref{tab: classification}, mean GenAI performance appeared to be weakest on tasks with a focus on \emph{Applications in new situations}, \emph{Implications, conjectures, and comparisons}, and \emph{Information transfer}. Mean GenAI attainment was approximately equal across the taxa of \emph{Factual knowledge} and \emph{Comprehension}, with the performance on the latter exceeding that on \emph{Routine use of procedures}. This indicates that GenAI attainment was not notably better on tasks that involved recall and standard application of learned techniques, though the strength of this conclusion is likely lessened by the omission of course materials from the prompts used in this study. Given the lack of course-specific knowledge provided with prompts, it is perhaps surprising that \cref{tab: classification} shows reduced performance in the taxa \emph{Applications in new situations} and \emph{Information transfer}. As questions in these categories required a baseline context for learned knowledge, this suggests that GenAI's effective context for learning is somehow in line with that of students or, perhaps more precisely, at least differs from the context of questions posed in the examination.

In addition to these distributions and taxon-level statistics, we provide further qualitative remarks and examples of observed themes in GenAI performance below.}

In procedural tasks, such as solving quadratic equations to determine eigenvalues, GenAI generally carried out the intermediate algebraic steps correctly. Yet, its solutions were frequently impeded by errors in the final stage of computation, where otherwise-correct manipulations produced incorrect numerical values. Notably, numerical accuracy accounted for only a small proportion of marks, so such minor errors did little to affect overall grades.

A similar pattern of behaviour was seen in structured, multi-part questions, designed to scaffold students towards the correct solution. For instance, on an example from number theory, GenAI consistently and correctly solved $7x\equiv1\mod40$ and, in a subsequent part, identified Fermat’s Little Theorem as helpful for solving $y^7 \equiv 12 \mod 41$. Nonetheless, the theorem was frequently misapplied and the earlier result ignored. Rather than use its earlier calculation, GenAI often attempted to repeat the step but made an error in formulating the problem, confusing the moduli 40 and 41, leading to few marks being awarded.

GenAI performance was \changed{also} somewhat unreliable when faced with `prove or disprove' questions, in which it was tasked with deciding if a statement was true or not and providing a proof or counterexample. For instance, it often struggled when asked whether or not the set of $2\times 2$ invertible matrices with positive determinant and trace forms a subgroup of $\mathrm{GL}_2(\mathbb{R})$. Its responses varied from incorrectly `proving' that it is a subgroup, to correctly identifying that the set is not a subgroup but then confidently providing an incorrect counterexample. On the other hand, GenAI was able to consistently produce standard examples of groups (e.g. those of order 8) whose elements have specific properties.

Another observed weak point in the performance of GenAI was its frequent inability to follow clear instructions in questions, such as being asked to rewrite a given formula without using specific notation. One example was seen in mathematical logic: when tasked with rewriting a formula whilst avoiding certain quantifiers or logical statements, it often failed to comply and simply continued to use the prohibited constructs. Despite this, GenAI's overall attainment in mathematical logic was strong.

Finally, we note that we used text-only prompts and accepted only text-based answers in this study. As such, one might expect questions involving sketches or graphs to limit GenAI performance. Perhaps surprisingly, we found no evidence of this. GenAI performed well on the two questions that asked respondents for a sketch, providing often comprehensible instructions that a competent user could translate into correct diagrams. Were GenAI given the opportunity to respond in image form, as is perhaps likely in a real-world setting, we expect that it may directly produce such diagrams, though we did not explore this.

\section{Discussion and conclusion}\label{sec: discussion}

Through a curriculum-wide evaluation of GenAI \changed{via 165 generated responses}, we have found that GenAI attainment in first-year undergraduate mathematics problems is on par with that of the student cohort \changed{(128 students)}, attaining a first-class result \changed{(see \cref{fig: programme})}. In other words, even without refined prompting or tailoring to the particular curriculum, the mastery demonstrated by GenAI tools is at the same level as that of students in examinations. We also found that GenAI responses, when aggregated across the curriculum, exhibited significantly lower variance than those of the students. Whilst this reduced variability is partially an expected feature of averaging \changed{statistically} independent responses, we observed limited overall variance, albeit with some slight module-to-module variance that is in line with the findings of \citet{udias2024chatgpt}. \changed{Given the synthetic nature of our evaluation, we highlight that the lack of variability may not transfer to real-world use of GenAI. For example, a student interacting with GenAI will plausibly introduce inter-question correlations, whilst varied prompting strategies may differently impact the format, accuracy, and variability of GenAI responses, likely in a way that is highly model dependent. We also found that variation in GenAI attainment was more evident when comparing different task types in \cref{tab: classification}, following the taxonomy of \citet{SmithTaxonomy}, though the broad distributions reported in \cref{fig: taxonomy} limited the suitability of summary statistics as descriptors. Nevertheless, we observed notable proficiency in routine applications of learned material and, somewhat surprisingly, similar performance in comprehension-based tasks.}

Our use of examinations as a proxy for course content allows us to draw an additional conclusion: the current style of questions used in invigilated examinations would not be fit for purpose if used as uninvigilated assessments, where GenAI should perhaps now be expected to be readily available to students. In other words, a first-class mark in such an assessment need not be a reliable indicator of a student's alignment with intended learning outcomes. Furthermore, assuming that examinations are intended in part to distinguish between students, the markedly low variance associated with GenAI responses suggests that this style of question would be wholly inappropriate for an unsupervised assessment. Whilst we do not expect that examinations created for invigilated assessments are widely thought to be suitable for use in uninvigilated contexts, informal discussions with colleagues from across institutions indicate that such use does occur. This is driven in part by heightened financial pressures on higher education institutions; this study provides empirical evidence against this practice.

It is important to highlight that there is considerable ambiguity in our primary conclusion of GenAI mastery. In particular, we have refrained from precisely defining mastery, save for using high examination marks as a proxy for it. There is a significant literature discussing the definition of mastery \cite{Nezhnov2015,NationalResearchCouncil2001,bloom1968learning,Simpson2023,Shearman2021}, accompanied by numerous studies evaluating whether or not in-person examinations are suited for assessing it \changed{in the context of a broader assessment diet \cite{BCollins2019,Iannone2022}}. GenAI's apparent mastery of first-year mathematics perhaps provides further motivation to question our collective interpretation of mastery, or at least to revisit how we assess university mathematics in practice. For instance, one might reasonably ask if our examination questions are testing skills that continue to be useful in the context of GenAI, given its potential for enabling cognitive offloading. In this respect, we argue that the problems posed in the evaluated examinations do retain considerable utility, in that we believe there is much to be gained by performing and understanding a process yourself even if it can be cognitively offloaded \cite{Risko2016}, but we recognise that there is likely to be wider debate on similar such questions. More generally, we expect GenAI to prompt further discourse on mathematics assessments and what it means to do a mathematics degree.

In future, rather than evaluating GenAI in the context of attainment, we hope to evaluate its suitability for enhancing student learning. This is a rapidly developing area of investigation that is demonstrating significant promise \cite{Kock2025,Contel2025,Kumar2023}, with a large proportion of students already making routine use of GenAI during otherwise independent study \cite{freeman2025student}. Whilst the accuracy of GenAI when used as a study partner may always prove to be a concern, \citet{Kumar2023} report that even engaging with erroneous GenAI outputs might be beneficial for learning. This adoption of GenAI risks bringing about a further source of wealth and digital inequality in higher education, with various tools offering paid services with purportedly more advanced capabilities \cite{McGee2025,Haddley2024}.

Given the rapid evolution of GenAI, our results are already expected to be outdated. As such, this case study serves not as a current benchmark, but as a lower bound on the curriculum-scale mathematical capabilities of GenAI in higher education. This is compounded by the constraints artificially imposed by this study, such as the absence of any opportunity to refine responses in collaboration with a user, or indeed the submission of course-specific lecture notes alongside an examination. Our conclusions, however, are unaffected; improvements in GenAI capability serve only to strengthen our remarks, with correspondingly greater impact on the pedagogical relevance of assessments and increased potential for heightening inequalities.

In summary, through a detailed, curriculum-wide evaluation of a popular GenAI tool, we have explored the proficiency of GenAI on a first-year undergraduate mathematics curriculum. Our findings demonstrate that GenAI is capable of exhibiting mastery of examination-style content at the level of a first-class degree. This was in line with the results of students, though GenAI was markedly more consistent when aggregated across the curriculum. Our observations underscore the need for the broad redesign of traditional assessments for use in open-book settings in the era of GenAI; the pressing challenge of how to meaningfully assess our students remains.

\section*{Ethics statement}
Ethical approval for the study was obtained from University College London's Humanities, Arts and Sciences Research Ethics Committee review board (reference HAS REC - 2025-0595-546). Student data was handled in accordance with GDPR. The Head of Department for the Department of Mathematics approved additional marking load for markers.

\section*{Funding}
BJW was supported by the Royal Commission for the Exhibition of 1851.

\appendix
\section{Module descriptions}\label{app: modules}
We briefly summarise the syllabi of the eight modules considered in this study, as delivered in academic year 2024--2025. Modules have been renamed for clarity.

\vspace{1em}

\noindent\textbf{Applied Mathematics: }An introduction to methods and tools used in mathematical modelling, including qualitative and analytic approaches to differential equations, stability, waves, oscillations, and dynamical systems.

\noindent\textbf{Calculus and Probability: }A primer in vectors, complex numbers, calculus, differential equations, and probability.

\noindent\textbf{Groups and Matrices: }An introduction to group theory and further linear algebra, including determinants and the principles of diagonalisation.

\noindent\textbf{Integration: }A rigorous theory of (Riemann) integration, along with concepts in sequences and series including uniform continuity and power series in the complex plane.

\noindent\textbf{Logic and Linear Algebra: }An introduction to algebra and discrete mathematics, primarily via linear equations. Includes the foundations of logic, set theory, functions and mappings, permutations, and fields.

\noindent\textbf{Mechanics and Dynamics: }A comprehensive introduction to the concepts of force, torque, momentum, angular momentum, and energy, in addition to the dynamics of point particles and celestial motion via vector methods. 

\noindent\textbf{Multivariable Calculus: }Introductions to programming and scientific computing (via Python) and the core methods and theorems of multivariable calculus.

\noindent\textbf{Real Analysis: }Starting only with the basic properties of real numbers, rigorous proofs of the main results in elementary differential calculus. Topics covered include sequences, series, continuity and differentiability of functions and the properties of the exponential function.

\newpage

\section{Question-by-question comparison}\label{app: questions}
\begin{figure}[H]
    \centering
    \includegraphics[width=0.9\linewidth]{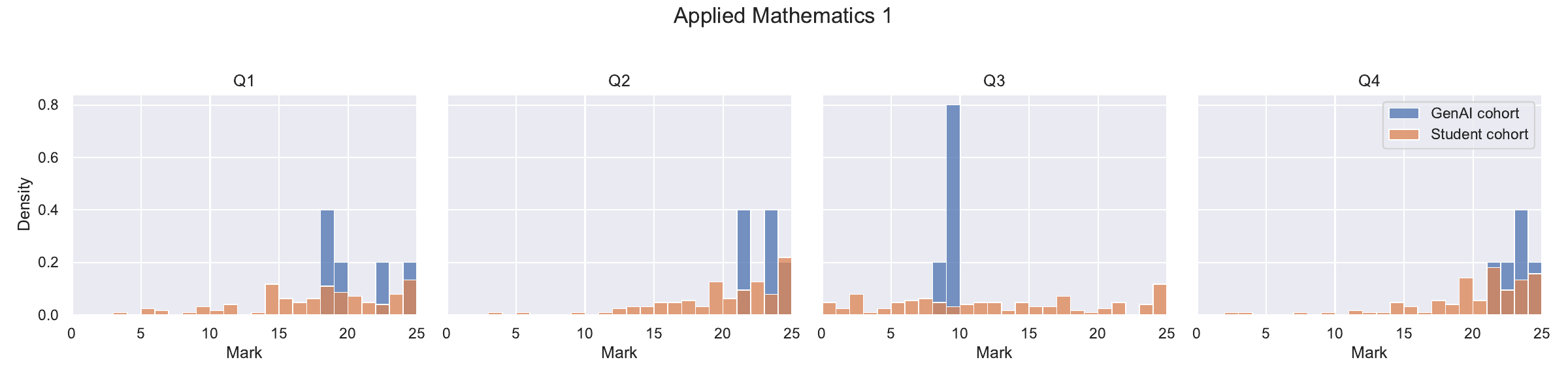}
    \includegraphics[width=0.9\linewidth]{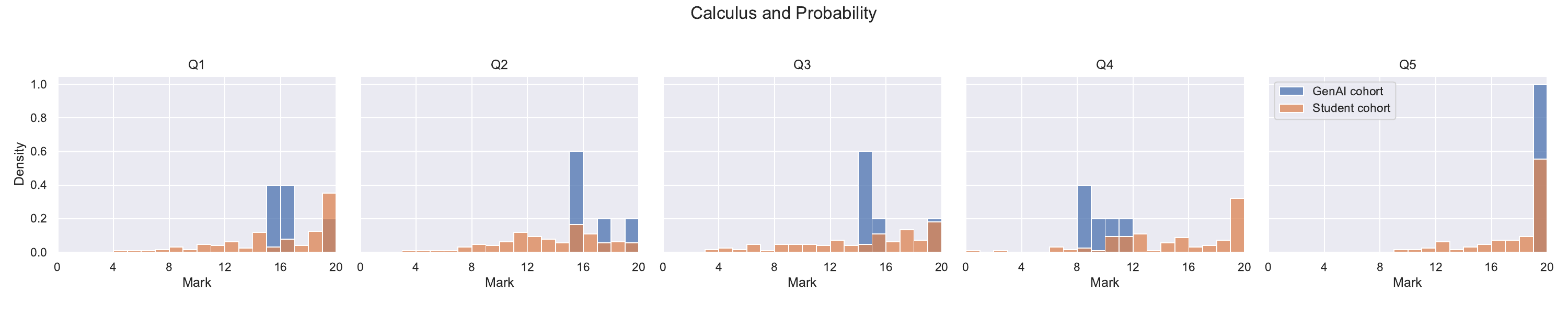}
    \includegraphics[width=0.9\linewidth]{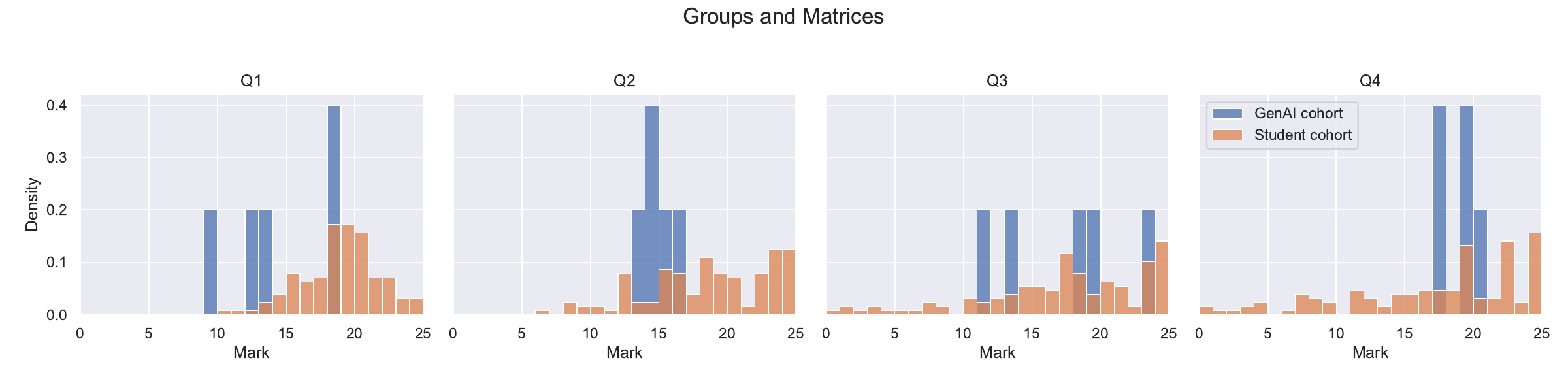}
    \includegraphics[width=0.9\linewidth]{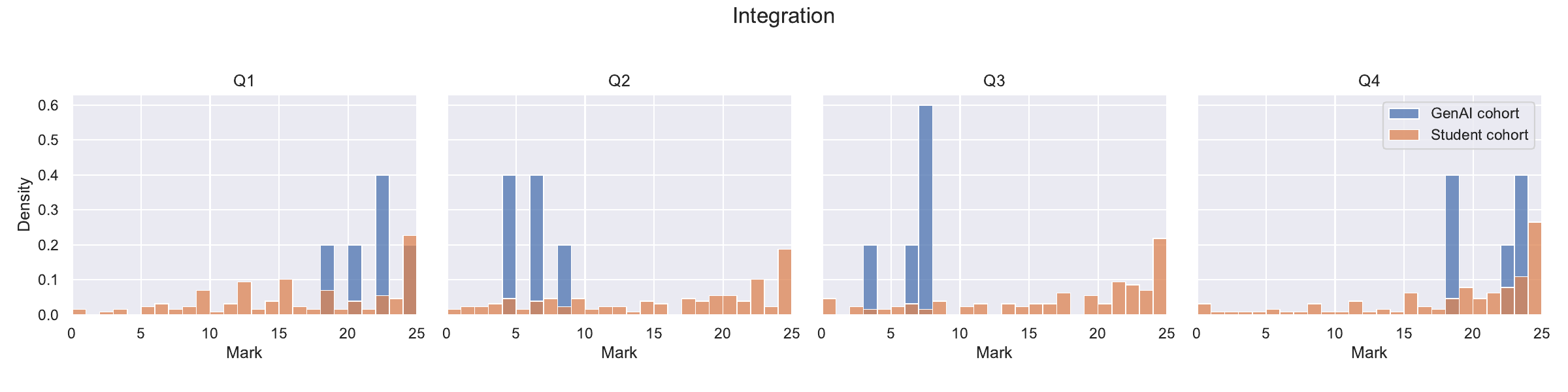}
    \caption{Question-by-question performance of GenAI compared to the student cohort in Applied Mathematics, Calculus and Probability, Groups and Matrices, and Integration. GenAI and student cohorts are shown in blue and orange, respectively.}
    \label{fig: questions 1}
\end{figure}

\begin{figure}[H]
    \centering
    \includegraphics[width=0.9\linewidth]{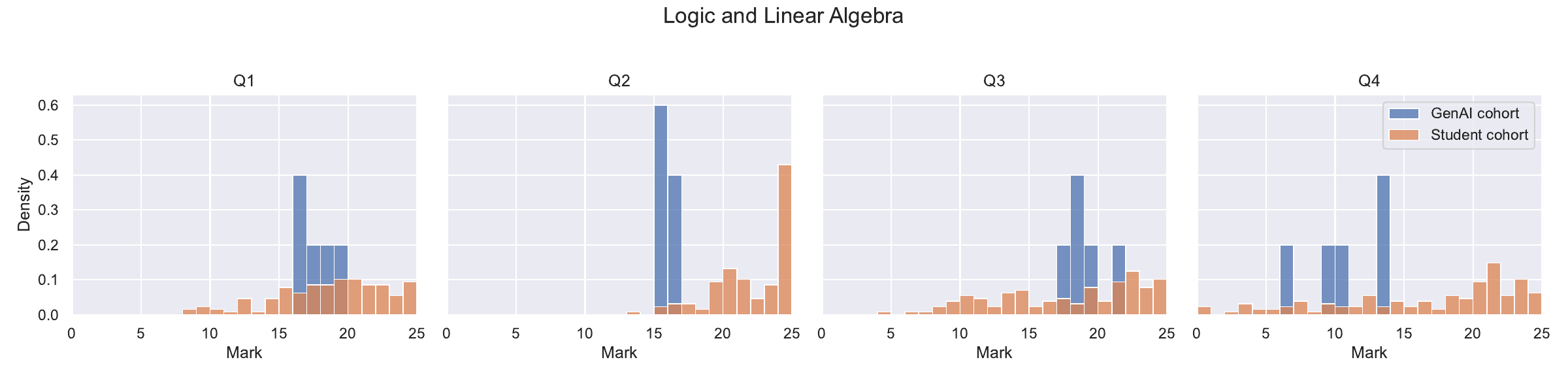}
    \includegraphics[width=0.9\linewidth]{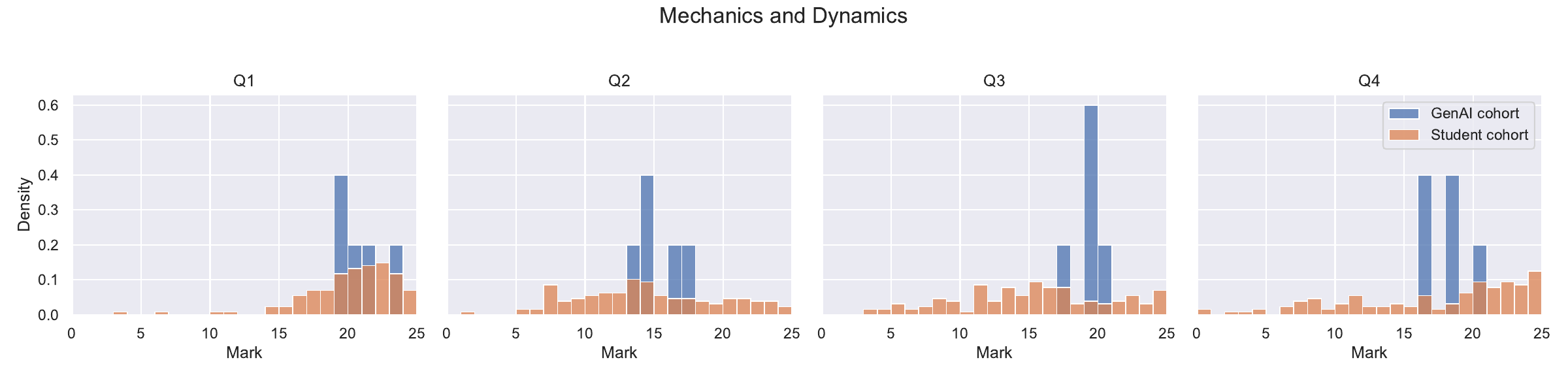}
    \includegraphics[width=0.9\linewidth]{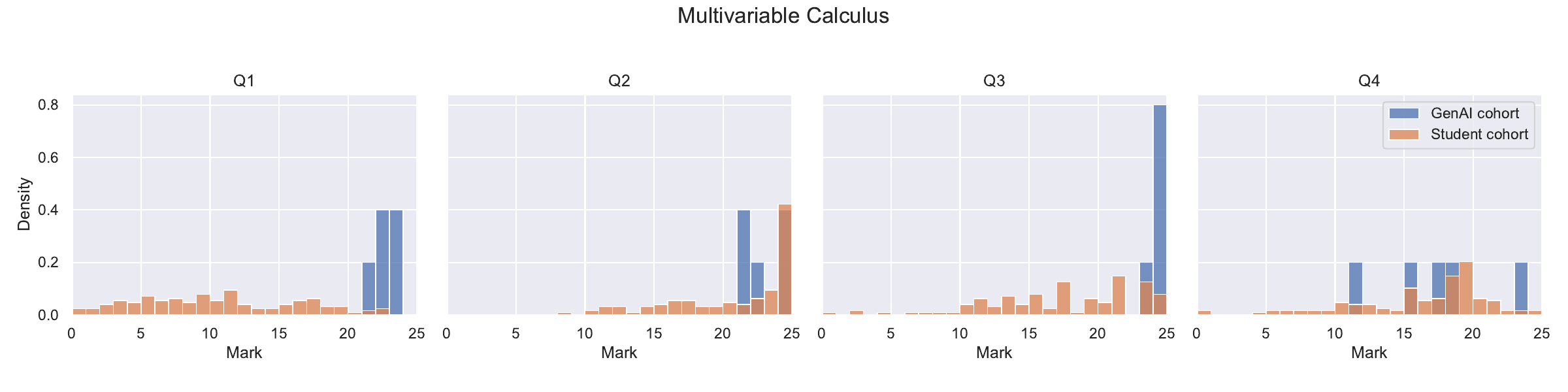}
    \includegraphics[width=0.9\linewidth]{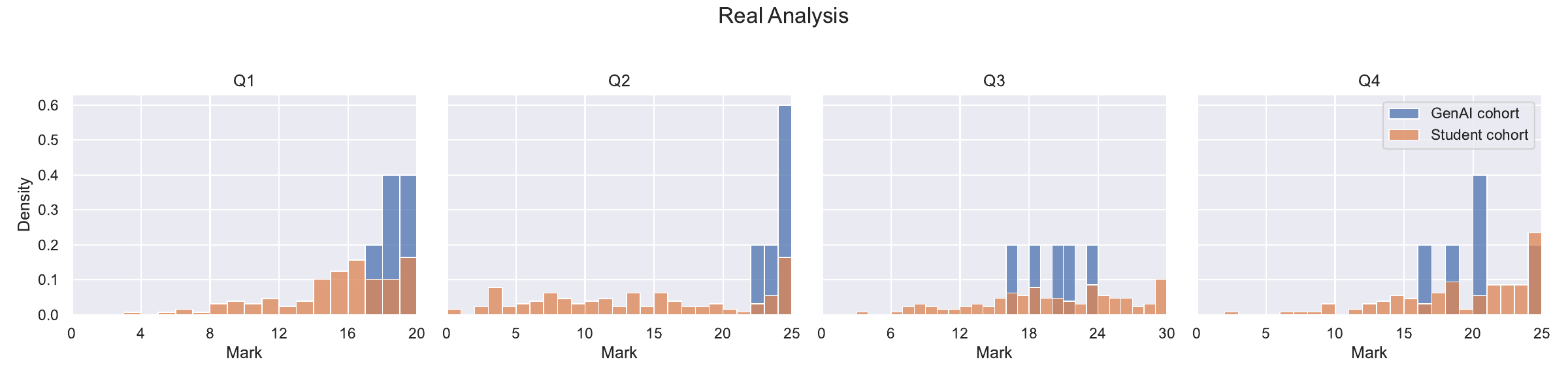}
    \caption{Question-by-question performance of GenAI compared to the student cohort in Logic and Linear Algebra, Mechanics and Dynamics, Multivariable Calculus, Real Analysis. GenAI and student cohorts are shown in blue and orange, respectively.}
    \label{fig: questions 2}
\end{figure}

We present the empirical distributions of GenAI and student cohort marks for each module, separated by question, in \cref{fig: questions 1,fig: questions 2}. Overall, GenAI appears to exhibit lower variance than the student cohort, though this depends strongly on the particular module and question. Similarly, average attainment is highly variable but, as shown in our curriculum-level analysis, the aggregated performance of GenAI was comparable to that of the student cohort.

\begin{figure}[H]
    \centering
    \includegraphics[width=1\linewidth]{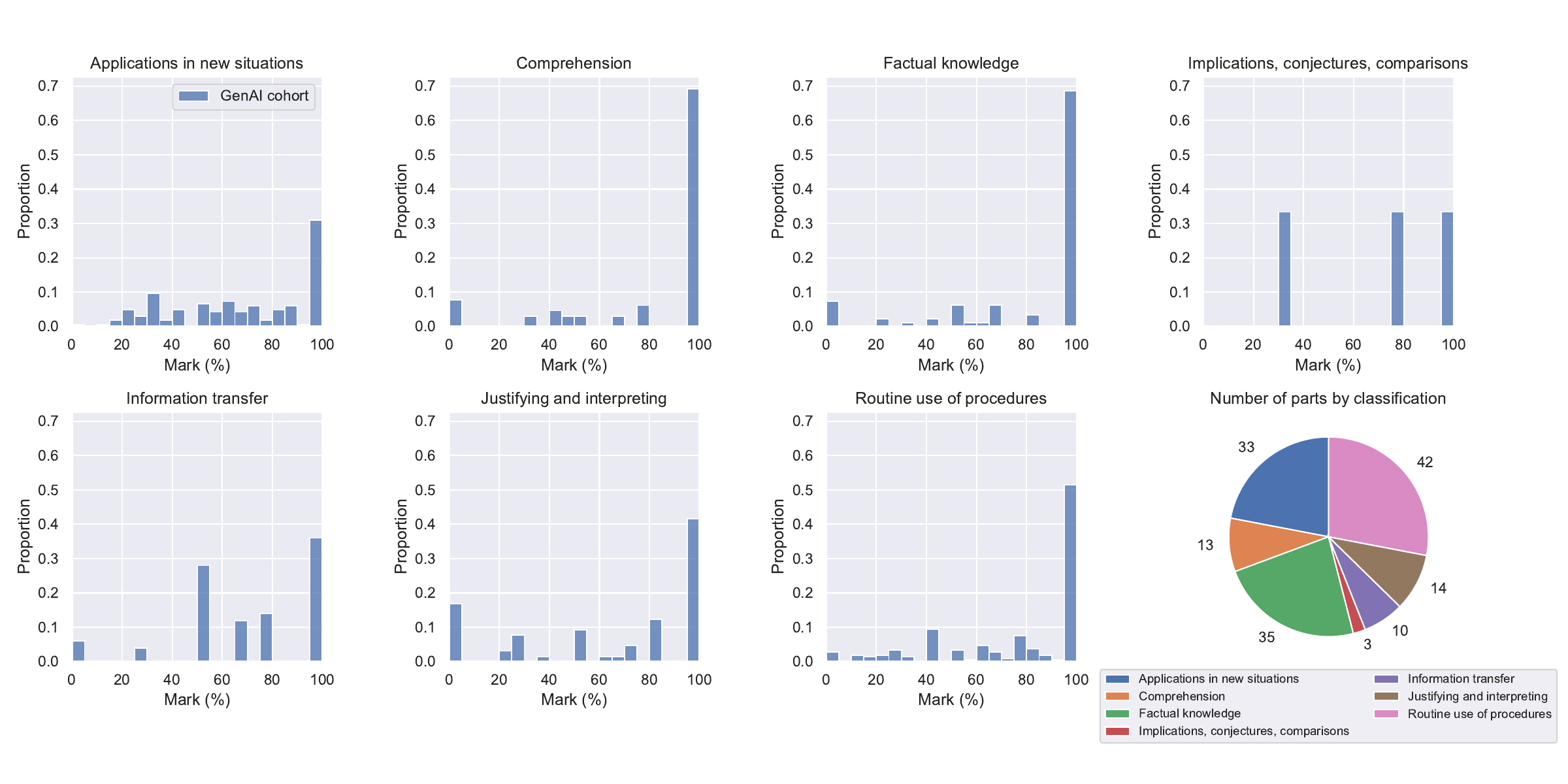}
    \caption{GenAI performance by classification of question, following the taxonomy of \citet{SmithTaxonomy}. Significant variability is present across all types of question, with significant clustering at 100\% attainment for \emph{Comprehension}, \emph{Factual knowledge}, and \emph{Routine use of procedures}. Performance is reported as percentage attainment on each question (or subpart, as necessary for unique classification). The lower right panel illustrates the numbers of questions/subparts in each taxon.}
    \label{fig: taxonomy}
\end{figure}

\changed{We also present the distributions of percentage attainment of GenAI on questions as classified following the taxonomy of \citet{SmithTaxonomy}, as shown in \cref{fig: taxonomy} and with questions split into subparts to enable classification. Whilst high spread is visible throughout, \emph{Comprehension}, \emph{Factual knowledge}, and \emph{Routine use of procedures} show significant peaks at 100\%. Further, 100\% attainment is the modal value for all types of question.}

\vspace{1em}\noindent
\textbf{Benjamin J. Walker} is a Lecturer in Applied Mathematics at University College London (UCL). He completed his DPhil in Mathematics at the University of Oxford. His research interests encompass mathematical biology, particularly the mechanics of microswimmers and the growth dynamics of biological tissues. He is a co-creator of \href{visualpde.com}{VisualPDE}, an interactive platform designed to facilitate the teaching and exploration of partial differential equations.

\vspace{1em}\noindent
\textbf{Nikoleta Kalaydzhieva} is a Lecturer (Teaching) in Pure Mathematics at UCL. She has been with UCL since her undergraduate studies, continuing through her PhD and into her current role. Her research interests lie in analytic number theory, focusing on problems related to multiple solutions of Pell's equation and continued fractions over function fields. She is chair of the Widening Participation Committee, Outreach Lead for the Department of Mathematics, and a Holgate Session Leader with the London Mathematical Society.

\vspace{1em}\noindent
\textbf{Beatriz Navarro Lameda} is a Lecturer (Teaching) in Pure Mathematics at UCL. She completed her PhD in Mathematics at the University of Toronto, Canada. Her research interests encompass mathematics education, probability theory, and dynamical systems. She is actively involved in the development and implementation of computer-aided assessment of mathematics and effective methods of teaching.

\vspace{1em}\noindent
\textbf{Ruth A. Reynolds} is a Lecturer (Teaching) in Pure Mathematics at UCL with a STACK focus. She completed her PhD in Noncommutative Algebra at the University of Edinburgh. Her research interests focus on the interactions between noncommutative ring theory and algebraic geometry, particularly in the study of idealizer subrings and their properties. In addition to noncommutative ring theory, Ruth has a long-standing interest in various aspects of mathematics education and e-assessment of mathematics using the STACK question-type.

\bibliographystyle{abbrvnat}
\bibliography{bib}

\end{document}